\documentclass[PRD,reprint,twocolumn,showpacs,
 nofootinbib, superscriptaddress]{revtex4}

\usepackage{graphicx}
\usepackage{dcolumn}
\usepackage{bm}

\newcommand{\beq}{\begin{equation}}
\newcommand{\eeq}{\end{equation}}
\newcommand{\beqa}{\begin{eqnarray}}
\newcommand{\eeqa}{\end{eqnarray}}

\begin{document}

\title{Signature of gravitational wave radiation in afterglows of short gamma-ray bursts?}
\affiliation{Key Laboratory of Dark Matter and Space Astronomy, Purple Mountain Observatory, Chinese Academy of Sciences, Nanjing 210008, China}
\affiliation{Chinese Center for Antarctic Astronomy, Purple Mountain Observatory, Chinese Academy of Sciences, Nanjing 210008, China }
\author{Yi-Zhong Fan}
\affiliation{Key Laboratory of Dark Matter and Space Astronomy, Purple Mountain Observatory, Chinese Academy of Sciences, Nanjing 210008, China}
\author{Xue-Feng Wu}
\affiliation{Chinese Center for Antarctic Astronomy, Purple Mountain Observatory, Chinese Academy of Sciences, Nanjing 210008, China }
\author{Da-Ming Wei}
\affiliation{Key Laboratory of Dark Matter and Space Astronomy, Purple Mountain Observatory, Chinese Academy of Sciences, Nanjing 210008, China}

\date{\today}

\begin{abstract}
Short Gamma-Ray Bursts (GRBs), brief intense emission of $\gamma-$rays characterized by a duration shorter than 2 seconds that are plausibly powered by the coalescence of binary neutron stars, are believed to be strong gravitational wave radiation (GWR) sources. The test of such a speculation has been thought to be  impossible until the performance of  the detectors like advanced LIGO. Recently there has been growing evidence for the formation of highly-magnetized neutron star (i.e., magnetar) in the double neutron star mergers.  In this work we re-examine the interpretation of the X-ray plateau followed by an abrupt decline detected in some short GRB afterglows within the supramassive magnetar model and find that the maximum gravitational mass of the non-rotating neutron stars is $\sim 2.3M_{\odot}$ and the observed duration of some X-ray plateaus are significantly shorter than that expected in the magnetic dipole radiation scenario, suggesting that the collapse of the supramassive magnetars has been considerably enhanced by the energy loss via GWR. Such a result demonstrates that the signature of GWR may have already existed in current electromagnetic data of short GRBs.
\end{abstract}


\maketitle
Thanks to the successful performance of the {\it Swift} satellite, our understanding of short GRBs, a kind of $\gamma-$ray outbursts with a duration less than two seconds \cite{Kouveliotou1993}, has been revolutionized \cite{Nakar2007}. At least for some short GRBs, the binary-neutron-star merger model \cite{Eichler1989} has been supported by their
host galaxy properties and by the non-association with
bright supernovae \cite{Gehrels2005,Leibler2010}. The
coalescence of double neutron stars inevitably produces an energetic burst of gravitational radiation, which is expected to be one of the most promising targets for current and the proposed future gravitational wave detectors \cite{Kochanek1993}. Moreover gravitational wave detection, in principle, is able to pin down the nature of the remnant of the merger, either a stellar mass black hole or a neutron star. Though the prospects are promising, so far direct detection of gravitational wave radiation (GWR) is not obtained yet and people are looking forward to the performance of detectors like advanced LIGO. However, we'll show in this short work that identifying GWR signature in current electromagnetic data of short GRBs is already possible.

In the relativistic simulations of double neutron star merger with a total gravitational mass $M_{\rm tot}\sim 2.6-2.8~M_\odot$,
initially a differentially-rotating heavy neutron star is formed \cite{Morrison2004,Shibata2005,Rosswog2007,Sekiguchi2011}. The magnetic braking and viscosity combine to drive the star to uniform rotation within a time $t_{\rm diff}\sim 0.1-1~{\rm s}$ if the surface magnetic field strength of the star reaches $10^{14}-10^{15}$ G \cite{Shapiro2000} and the magnetic activity of the initial differentially-rotating neutron star is likely able to drive brief energetic $\gamma-$ray outburst and thus account for the short GRB \cite{Rosswog2007}. The uniform rotation period of the heavy neutron star is $P_0\sim 1$ ms  (\cite{Sekiguchi2011}, following \cite{Lee01} one can also show analytically that it is the case). The rapid rotation can enhance the maximum gravitational mass ($M_{\rm max}$) by a factor of $\sim 0.05(P_0/1~{\rm ms})^{-2}$ and thus help to make the nascent heavy neutron star stable, i.e., in the presence of rotation the allowed gravitational mass is $M_{\rm max,r}\approx [1+0.05(P_0/1~{\rm ms})^{-2}]M_{\rm max}$ \cite{Friedman1986}. For the merger-formed suprammasive neutron star that is supported against collapse by uniform rotation, the stability condition reads $M_{\rm max,r}+\alpha (M_{\rm max,r}/1~{M_\odot})^{2}+m_{\rm loss}-2[\alpha (M_{\rm ns}/1~M_\odot)^2+M_{\rm ns}]>0$ (i.e., the baryonic mass of the newborn remnant should be smaller than the allowed maximum one), where $\alpha\approx 0.084~M_\odot$, $m_{\rm loss}$ is the mass loss during the merger process and $M_{\rm tot}=2M_{\rm ns}$ has been assumed for simplicity \cite{Rosswog2007}. As a conservative estimate, we take $m_{\rm loss}=0$. It is straightforward to show that for sufficient stiff equation of states yielding $M_{\rm max}\gtrsim 2.3~M_\odot$, the merger of double neutron stars with a total gravitational mass $M_{\rm tot}\lesssim 2.6~M_\odot$ can produce a supramassive neutron star with  $P_0 \sim 1$ ms, which survives until a good fraction of its rotational energy has been lost. The newborn neutron star is likely highly-magnetized, i.e., it is a  millisecond magnetar \cite{Gao2006,Duncan1992,Rosswog2007}. On the one hand, PSR J0348+0432 has an accurately measured gravitational mass $2.01\pm 0.04~M_\odot$ \cite{Demorest2010} and gravitational masses in the range of $2.4-2.7~M_\odot$ have been reported for a few pulsars though the uncertainties are still large \cite{Lattimer2012}. On the other hand, among the ten Galactic neutron star binary systems detected so far,  five have an $M_{\rm tot}\approx 2.6~M_\odot$ \cite{Lattimer2012}. These two sets of observational facts strongly suggest that supramassive magnetars are likely the remnants of a good fraction of double neutron star mergers.
The internal energy dissipation of the Poynting-flux dominated outflow launched by the newly-formed millisecond magnetar before its collapse may power a high energy transient which accounts for the peculiar X-ray emission (i.e., unexpected in the external forward shock model) following short bursts \cite{Gao2006}, in particular the X-ray plateau lasting $\sim 100$ s followed by an abrupt cease \cite{Rowlinson2010b}.

In a very recent systematic analysis of the X-ray afterglow emission of short GRBs, Rowlinson et al. \cite{Rowlinson2013} identified such a kind of supramassive magnetar signature in about half of events. As usual, these authors attributed the abrupt termination of the X-ray plateau to the collapse of the supramassive magnetar and then estimated the physical parameters, including the initial rotation period $P_0$ and the strength of dipole magnetic field $B_{\rm \perp}$ with the widely-adopted assumptions that the duration (luminosity in a very wide energy range of $1-10^{4}$ keV, $L_{_{\rm plat}}$) of the plateau is the spin-down timescale (dipole radiation luminosity) of the magnetar, i.e.,
\begin{eqnarray}
t_{_{\rm plat}} &\approx & (1+z)\tau_{\rm dip} \nonumber\\
&\approx & 4\times 10^{3}~{\rm s}~(1+z)({I\over 10^{45.3}{\rm g~cm^{2}}}) ({P_{0}\over 1{\rm ms}})^{2} \nonumber\\
&& ({R_{\rm s}\over 10^{6}{\rm cm}})^{-6} ({B_{\rm \perp}\over 10^{15}{\rm G}})^{-2},
\label{eq:t_dip}
\end{eqnarray}
\begin{equation}
L_{_{\rm plat}}\approx 10^{49}~{\rm erg~s^{-1}}~\eta({R_{\rm s}\over 10^{6}{\rm cm}})^6({B_{\rm \perp}\over 10^{15}{\rm G}})^{2}({P_0\over 1{\rm ms}})^{-4},
\label{eq:L_x}
\end{equation}
where $R_{\rm s}$ ($B_{\rm s}$) is the radius (surface polar magnetic field strength) of the star, $B_{\rm \perp}=B_{\rm s} \sin \alpha$ and $\alpha$ is the angle between the rotational and dipole axes \cite{Shapiro1983}, $z$ is the redshift of the GRB, and $\eta$ is the efficiency of converting the magnetar wind energy into late prompt radiation and has been taken to be $\sim 1$ in \cite{Rowlinson2010b,Rowlinson2013}. The inferred $B_{\rm \perp}$ are in the range $10^{15}-10^{17}$ Gauss, in agreement with the magnetar hypothesis. The inferred $P_0$ for most events however are $\gg 1$ ms, at odds with the double neutron star merger model in which $P_0 \sim 1$ ms is expected. Moreover, the inferred rotation period $P_0\gg 1$ ms is too slow to support a supramassive compact object \cite{Friedman1986}, for which it is improper to assume that the remnant collapse is triggered by the loss of a large amount of rotation energy (Consequently, one can not use eq.(\ref{eq:t_dip}) to estimate the physical parameters of the newborn compact object any longer). For illustration we discuss GRB 080905A and GRB 090515A, two nearest short events with magnetar central engine signature in the X-ray lightcurves. With the X-ray plateau of GRB 090515A at a redshift $z=0.405$ \cite{Leibler2010}, $P_0$ is inferred to be $\sim 10$ ms \cite{Rowlinson2010b}. For 080905A at a redshift $z=0.122$ \cite{Rowlinson2010a}, $P_0$ is estimated to be $\sim 5$ ms \cite{Rowlinson2013}. As already mentioned, such spin periods are considerably longer than that expected in the double neutron star merger model. The current situation can be re-expressed more straightforwardly as that the total energy released in the X-ray plateau phase is much smaller than that expected in the merger-formed magnetar model ($\sim 2\pi^2 I/P_0^2 \sim 10^{52}~{\rm erg}$) \footnote{For GRB 080905A, the prompt emission has an isotropic $\gamma-$ray energy $\sim 6\times 10^{50}$ erg \cite{Rowlinson2013}, which is much smaller than the rotation energy of the newborn magnetar with $P_0 \sim 1$ ms. The speculation that much more energy has been lost via hard $\gamma$-rays ($>10$ MeV) is not supported by current Fermi-LAT observations of short GRBs, either.}. Such a puzzle might be solved in two ways. One is that $\eta$ is as low as $\sim 0.01$, then the intrinsic $P_0$ can be $\sim 1$ ms since
$L_{_{\rm plat}}t_{_{\rm plat}}/(1+z) \approx 4\times 10^{52}~{\rm erg}~\eta ({I/10^{45.3}{\rm g~cm^{2}}})({P_{0}/ 1{\rm ms}})^{-2}$, which yields $P_0 \propto \eta^{1/2}$ for given $L_{_{\rm plat}}$ and $t_{_{\rm plat}}/(1+z)$.
Such a small $\eta$ however may be less likely since usually the magnetic energy dissipation process is expected to produce (late) prompt emission at a moderate/high efficiency $\sim 0.5$ \cite{Drenkhahn2002}, that is why in the literature (e.g., \cite{Rowlinson2010b,Rowlinson2013}) people adopted $\eta \sim 1$. Moreover in the low $\eta$ case, the non-dissipated magnetar wind will inject into the blast wave of the GRB outflow, drive energetic forward shock and give rise to bright afterglow emission \cite{Dai1998}. GRB 090515A likely lies outside of its host galaxy \cite{Leibler2010} and one might be able to attribute the very dim optical emission to the extremely low circum-burst medium. However for GRB 080905A such an argument no longer applies since this burst offsets from the host galaxy centre
by $\sim 18.5$ kpc but still in the northern spiral arm which exhibits an older stellar population than the
southern arm \cite{Rowlinson2010b}. Hence we expect a normal circum-burst medium, with a kinetic energy $> 10^{52}$ erg and a very low redshift $z=0.122$ the forward shock emission should be extremely bright. The observations are not in support of such a speculation. Actually at $t\sim 8.5$ hours after the trigger, the R-band emission was at the magnitude of $24.04\pm 0.47$. At late times it faded continually and was not detected any longer at $t>36$ hours after the trigger \cite{Rowlinson2010a}. Shifting the burst to $z=1$, at the observer's time $\sim 15.2$ hours after the trigger the R-band emission should be $\sim 29^{\rm th}$ magnitude. However, as shown in Fig.15 of \cite{Nicuesa2012}, the bursts with a typical kinetic energy $\sim 10^{52}$ erg (most are long events) are far brighter. We therefore conclude that for some events the small $\eta$ argument is disfavored  (this conclusion is also favored by the rare detection of radio afterglow for short GRBs) and then turn to the other possibility that most of the rotational energy was carried away via the strong gravitational wave radiation.
Rotating magnetar with a triaxial shape has a time varying quadrupole moment and hence radiates gravitational waves at a frequency $f=2/P$ \cite{Shapiro1983,Howell2011}. A magnetar loses rotational energy through magnetic dipole torques and gravitational wave radiation emission \cite{Shapiro1983}
\[
-dE_{\rm rot}/dt=\pi^{4}R_{\rm s}^6B_{\rm \perp}^{2}f^{4}/6c^{3}+32\pi^{6}GI_{\rm zz}^{2}\epsilon^{2}f^{6}/5c^{5},
\]
where $\epsilon=2(I_{\rm xx}-I_{\rm yy})/(I_{\rm xx}+I_{\rm yy})$ is the ellipticity in terms of the principal moments of inertia (i.e., $I$). We would like to point out that the gravitational wave radiation of magnetars formed in collapsars driving long GRBs has been extensively investigated \cite{Corsi2009,Howell2011} but these works focus on the detectability by the future detectors and no gravitational radiation signature already existed in the X-ray afterglow emission has been suggested possibly because the spin periods of such a kind of magnetars are hard to reliably estimate.

For
\[
\epsilon<1.5\times 10^{-3} ({I_{\rm zz}\over 10^{45.3}~{\rm g~cm^{2}}})^{-1} ({P_{0}\over 1~{\rm ms}})({R_{\rm s}\over 10^{6}~{\rm cm}})^{3} ({B_{\rm \perp}\over 10^{15}~{\rm G}}),
\]
the rotational energy loses mainly through the dipole radiation and the spin down timescale is estimated by eq.(\ref{eq:t_dip})
otherwise the rotational energy loses mainly via the gravitational wave radiation and the spin down timescale is given by
\[
\tau_{_{\rm GW}} \approx 90~{\rm s}~({I_{\rm zz}\over 10^{45.3}{\rm g~cm^{2}}})^{-1}({P_{0}\over 1{\rm ms}})^{4} ({\epsilon\over 0.01})^{-2}.
\]
For $t> (1+z)\min\{\tau_{\rm dip},~\tau_{_{\rm GW}}\}$, a considerable fraction of rotation energy has lost and the decelerated magnetar likely collapses. The duration of the plateau can be estimated as $t_{\rm plat} \approx (1+z)\min\{\tau_{\rm dip},~\tau_{_{\rm GW}}\}$ and the luminosity is given by eq.(\ref{eq:L_x}).
Hence the puzzle arising in interpreting the X-ray plateau data of GRB 080905A and GRB 090915A within the merger-formed supramassive magnetar model can be solved as long as $t_{\rm plat}\approx (1+z)\tau_{_{\rm GW}}$, requiring
\begin{equation}
\epsilon \approx 0.0095~({I_{\rm zz}\over 10^{45.3}{\rm g~cm^{2}}})^{-1/2}({P_{0}\over 1{\rm ms}})^{2}[{t_{\rm plat}\over (1+z)100{\rm s}}] ^{-1/2},
\label{eq:epsilon_constraint}
\end{equation}
which is considerably larger than
the maximum elastic quadrupole deformation of conventional neutron stars, but comparable to the
upper limit derived for crystalline colour-superconducting quark
matter \cite{Lin2007}. Interestingly the required sizable ellipticity may be attributed to the deformation caused by the strong  magnetic field of the magnetar (see \cite{Howell2011} for a summary and for other possibilities). If the internal magnetic field
is purely poloidal and matches the dipolar field in the exterior, the
ellipticity may be as large as $\sim 0.01$ for $B_{\rm \perp} \sim 5\times 10^{15}$ G \cite{Bonazzola1996}. As shown in eq.(\ref{eq:L_x}), such a large $B_{\rm \perp}$  is too large in the moderate/high $\eta$ case but possible in the low $\eta~ (\sim 0.01)$ case.  Alternatively the
internal magnetic field $B_{\rm t}$ of the magnetar can be dominated by a toroidal
component. The enormous energy liberated in the 2004 December 27 giant flare from SGR 1806-20 (the central engine is widely believed to be an old magnetar) together with the likely recurrence time of such events suggests $B_{\rm t} \sim 10^{16}$ G \cite{Stella2005}. For the newborn magnetars,  $B_{\rm t} \sim 10^{17}$ G is likely, which may have decayed by one order of magnitude in $\sim 10^{4}$ years  \cite{Duncan1992,Stella2005}. Such super-strong toroidal fields could induce
a prolate deformation with the ellipticity
$\epsilon \sim 0.016 (B_{\rm t}/10^{17}~{\rm G})^{2}$  \cite{Usov1992,Stella2005},
matching that needed in eq.(\ref{eq:epsilon_constraint}).

In our scenario, the gravitational wave radiation has a frequency $f\approx 2000$ Hz and the energy spectrum is $dE_{\rm gw}/df\approx \pi^{2}I_{\rm zz}f$, hence the characteristic gravitational wave amplitude can be estimated as \cite{Howell2011,Corsi2009}
\begin{eqnarray}
h_{\rm c}&=&{1\over \pi D}\sqrt{{5G \over 2c^{3}}{dE_{\rm gw}\over df}}\nonumber\\
&=& 5.1\times 10^{-22}({100~{\rm Mpc} \over D})({I_{\rm zz}\over 10^{45.3}~{\rm g~cm^2}})^{1/2}({P_0\over 1~{\rm ms}})^{-1},
\end{eqnarray}
which is below (but comparable to) the advanced-LIGO noise curve $h_{\rm rms}=[fS_{\rm h}(f)]^{1/2}$ \cite{Harry2010} (where $S_{\rm h}(f)$ is the power spectral density of the detector noise) at such a high frequency, where $D$ is the distance of the burst to the earth. The proposed Einstein Telescope with an expected $h_{\rm rms}\approx 5\times 10^{-23}$ at $f=2000~{\rm Hz}$ \cite{Punturo2010} may be able to detect the gravitational radiation signal discussed in this work. Interestingly, with the Einstein Telescope the remanent (either a black hole or a supramassive neutron star) formed in double neutron star merger at a distance $D\leq 1$ Gpc may be directly pinned down \cite{Hotokezaka2011} and the collapse process might be identified too \cite{Novak1998}. Though the magnetar gravitational wave signal discussed in this work may still be undetectable if $D>100$ Mpc, one can however test our model indirectly by monitoring the electromagnetic counterpart to estimate the total energy injected into the medium since one of our main predictions is that such an energy should be much smaller than the kinetic energy of a supramassive magnetar.\\

\acknowledgments  This work was supported in part by 973 Programme of China under grants 2013CB837000 and 2009CB824800, National Natural Science of China under grants 11073057 and 11273063, and the Foundation for
Distinguished Young Scholars of Jiangsu Province, China (No. BK2012047).  YZF and XFW are also supported by the 100
Talents programme of Chinese Academy of Sciences.

\end{document}